**Intrinsic to extrinsic phonon lifetime transition in a GaAs-AlAs superlattice**

F Hofmann[1,2]*, J Garg[3], AA Maznev[2], A Jandl[4], MT Bulsara[4], EA Fitzgerald[4], G Chen[3], KA Nelson[2]

[1] Department of Engineering Science, University of Oxford, Parks Road, Oxford, OX1 3PJ, UK

[2] Department of Chemistry, Massachusetts Institute of Technology, Cambridge, MA 02139, USA

[3] Department of Mechanical Engineering, Massachusetts Institute of Technology, Cambridge, MA 02139, USA.

[4] Department of Materials Science and Engineering, Massachusetts Institute of Technology, Cambridge, MA 02139, USA.

* *Corresponding author: felix.hofmann@eng.ox.ac.uk*

**Abstract**

We have measured the lifetimes of two zone-center longitudinal acoustic phonon modes, at 320 GHz and 640 GHz respectively, in a 14 nm GaAs / 2 nm AlAs superlattice structure. By comparing measurements at 296 K and 79 K we separate the intrinsic contribution to phonon lifetime determined by phonon-phonon scattering from the extrinsic contribution due to defects and interface roughness. At 296 K, the 320-GHz phonon lifetime has approximately equal contributions from intrinsic and extrinsic scattering, whilst at 640 GHz it is dominated by extrinsic effects. These measurements are compared with first-principles lattice dynamics calculations of intrinsic and extrinsic scattering rates in the superlattice. The calculated room-temperature intrinsic lifetime of longitudinal phonons at 320 GHz is in agreement with the experimentally measured value of 0.9 ns. The model correctly predicts the transition from predominantly intrinsic to predominantly extrinsic scattering; however the predicted transition occurs at higher frequencies. Our analysis indicates that the "interfacial atomic disorder" model is not entirely adequate and that the observed frequency dependence of the extrinsic scattering rate is likely to be determined by a finite correlation length of interface roughness.

**I. Introduction**

Semiconductor superlattices (SLs) are excellent model systems for studying phonons in nanostructured materials.[1] The GaAs-AlAs system allows fabrication of high-quality lattice-matched SLs and has attracted attention in the context of both heat transport [2-6] and coherent phonon studies.[7-11] While acoustic phonon dispersion in SLs is well-understood,[8, 12, 13] the same cannot be said about phonon transport properties. In particular, quantitative understanding of room-temperature thermal conductivity of SLs is still lacking [1] despite recent progress in both theory [5, 14-17] and experiment.[18] Compared to bulk GaAs, calculations of thermal transport in defect-free GaAs-AlAs SLs with perfect interfaces yield a smaller reduction in thermal conductivity [16] than experimental observations.[2] Thus it is believed [16, 17] that "extrinsic" processes such as defect scattering and interface roughness govern the observed thermal conductivity. In that case one would expect thermal conductivity to decrease when the temperature is lowered,[17] as extrinsic scattering rates remain constant but the phonon population decreases. However experimental results actually showed a slight increase in thermal conductivity of GaAs-AlAs SLs when the temperature was decreased from 370 K to 100 K.[2] Thus it appears likely that thermal conductivity of SLs is controlled by the interplay of both intrinsic and extrinsic scattering processes.

The key issues for understanding thermal transport in SLs and other nanostructured materials are the relative roles of intrinsic and extrinsic scattering in determining the frequency-dependence of phonon lifetimes. Phonon lifetimes cannot be obtained from thermal conductivity measurements that yield information integrated over the entire phonon spectrum. Laser excitation of coherent phonons allows the direct measurement of specific phonon modes. Unfortunately progress in harnessing coherent phonons for studying phonon lifetimes at high frequencies has been slow. In bulk single crystals, room-temperature coherent phonon lifetimes have been measured at frequencies up to 100 GHz in Si [19] and up to 56 GHz in GaAs.[20] Data on coherent phonon lifetimes in SLs are even scarcer. In one study a room-temperature lifetime of 250 ps for 0.6 THz longitudinal acoustic (LA) phonons in a GaAs-AlAs

SL with 8 nm period was reported,[21] although the dominant scattering mechanism was not determined. Lifetimes of phonon nanocavity modes, which are related but not equivalent to SL modes, were reported in refs. 22 and 23. The apparent lifetime of 50 ps measured at 1 THz was found to be predominantly extrinsic [23] and ascribed to inhomogeneity caused by layer thickness variations, i.e. to inhomogeneous dephasing that masked the true lifetime as discussed further below. We recently reported the lifetime of 330 GHz LA phonons in an 8 / 8 nm GaAs/AlAs SL structure.[24] By considering measurements at 296 K and 79 K we were able to separate intrinsic and extrinsic scattering rates and deduced an intrinsic lifetime of 0.95 ns, whereas the extrinsic lifetime was 0.69 ns. However the fact that only one frequency could be accessed limited the value of this information. The frequency of laser-excited coherent phonons generated in a SL can be varied by changing the SL period. But since the extrinsic contribution to phonon lifetime may be affected by variations in the fabrication process, it is desirable to measure the frequency dependence of phonon lifetimes in a single sample. Fortunately, multiple (nearly harmonic) zone-center modes are typically observed on SLs with asymmetric layer thicknesses.[25]

In the present study, we aim to gain insight into scattering mechanisms and the frequency dependence of phonon lifetimes by measuring two LA phonon modes (320 and 640 GHz) in a 14 / 2 nm GaAs/AlAs SL structure. By comparing measurements at 296 K and at 79 K we separate extrinsic and intrinsic scattering contributions. We find that a transition from predominantly intrinsic to predominantly extrinsic scattering takes place in the sub-THz range. The results are compared to first-principles lattice dynamics-based calculations of phonon lifetimes in the SL, that use interatomic force constants derived from density-functional perturbation theory (DFPT). We calculated both three-phonon and interfacial scattering rates, using an atomic disorder model to capture the latter. We will see that the calculations predict anharmonic scattering rates well, whilst our interface scattering model requires further refinement.

## II. Experiments

The SL sample consisted of 219 periods of nominally 14 nm GaAs [001] and 2 nm AlAs [001] layer pairs (period $d$ = 16 nm) on a GaAs substrate. It was grown by metal-organic chemical vapor deposition (MOCVD) at a temperature of 750° C. A 500 nm homoepitaxial layer of GaAs was initially deposited, followed by the SL structure.

Coherent phonons were excited by pumping the structure with 300 fs (FWHM) pulses of 784 nm laser light. At this wavelength the pump photon energy (1.58 eV) lies above the bandgap of GaAs (1.42 eV), but below that of AlAs (2.15 eV). Hence, whilst AlAs appears almost transparent, excited carriers generated in GaAs lead to mechanical stress via the deformation potential.[26] This periodic transient mechanical stress launches the coherent phonons. They are easily detected, as mechanical strain couples to the optical constants, causing changes in the reflectivity of the structure when monitored with a time-delayed probe pulse. A schematic illustration of the experiment is shown in Fig. 1 (a). The output from an amplified Ti:Sapphire system (784 nm wavelength, 300 fs pulse duration, 250 kHz repetition rate) was split into pump and time-delayed probe beams. The pump pulses were modulated at 93 kHz by an acousto-optic modulator to facilitate lock-in detection, and focused to a 100 μm diameter (1/e intensity) spot on the sample. Pump pulse energy of 70 nJ was used for measurements at 296 K, whilst at 79 K the pump energy was 180 nJ. The time-delayed probe beam (pulse energy 25 nJ) was focused to a 30 μm diameter spot (1/e level) at the center of the pumped region. After reflection from the sample it was detected by a photodiode whose output was fed into a lock-in amplifier. Measurements were averaged over several hundred scans of the optical delay line with an average data collection time of ~20 hours per experiment. To allow measurements at different temperatures the sample was mounted in a nitrogen-cooled cryostat.

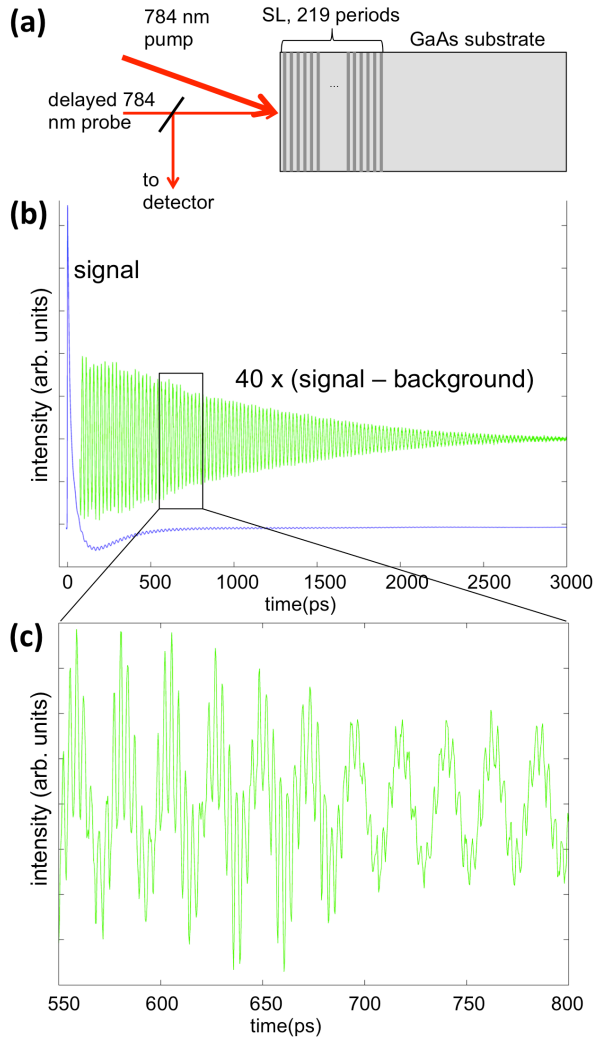

*Figure 1: a) Schematic illustration of the experimental arrangement and the SL structure (not to scale). b) Experimentally recorded sample reflectivity signal at 79 K. Coherent phonon oscillations after subtraction of background (inset). c) Derivative of phonon oscillation signal in a time window from 550 to 800 ps.*

### III. Data interpretation and results

A time trace of the reflected intensity signal recorded at 79 K is shown in Fig. 1 (b). It shows a large electronic/thermal response with an initial high peak, followed by slow dynamics. Superimposed on this are small acoustic oscillations that can be isolated by removing the slow background signal. The isolated acoustic signal is inset in Fig. 1 (b). Fig. 1 (c) shows a magnified view of the derivative of the coherent phonon signal in a time window from 550 ps to 800 ps. Oscillations at different frequencies are clearly visible. Fourier transforms of the acoustic oscillations, evaluated in three time windows, $T_1$ (50 ps to 350 ps), $T_2$ (1050 ps to 1350 ps) and $T_3$ (2050 ps to 2350 ps), are shown in Fig 2 (a). In the $T_1$ spectrum a number of peaks can be seen, namely the first zone-center mode (ZC1) at 319 GHz and its "satellites" (BS2 and BS3) at 281 GHz and 369 GHz respectively, and the second zone-center mode (ZC2) at 643 GHz with satellites (BS4 and BS5) at 605 GHz and 695 GHz respectively.

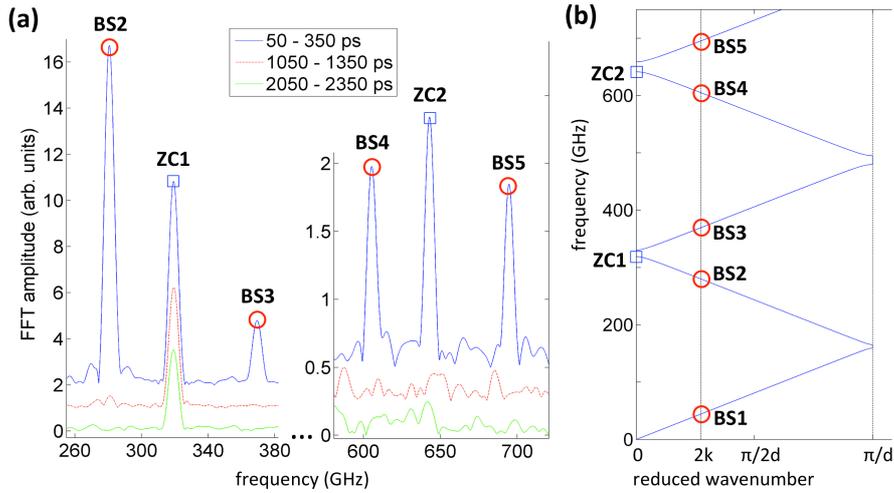

*Figure 2: a) Fourier amplitude spectra of coherent phonon oscillations at 79 K computed in three different time windows, top to bottom: $T_1$ (50 – 350 ps), $T_2$ (1050 – 1350 ps) and $T_3$ (2050 – 2350 ps). b) Calculated dispersion curve for LA phonons within the first mini-Brillouin zone at 79 K. Superimposed are the experimentally measured frequencies.*

In Fig. 2 (b) the observed frequencies are compared with the computed dispersion relation of the folded LA phonon branch in the mini Brillouin zone.[12] We used acoustic velocities of 4719 ms$^{-1}$ and 5718 ms$^{-1}$ for GaAs and AlAs respectively.[27, 28] To match the experimentally observed frequencies at 296 K the GaAs layer thickness was adjusted to 12.9 nm, whilst the AlAs layer thickness was maintained at the nominal 2.0 nm. The ZC1 and ZC2 peaks correspond to the symmetric, zero reduced wavenumber zone-center modes of the SL and lie at the bottom of small bandgaps (Fig. 2 (b)). The antisymmetric zone-centered modes at the tops of the bandgaps are not excited in our measurements. The satellite peaks, BS1 - BS5, are due to modes with wave number $q = 2k$ where $k$ is the wavenumber of the probe light in the medium.[10] Their positions are marked on the dispersion curves. The BS1 peak at 44.1 GHz is present in the spectrum, but is not shown in Fig. 2 (a) for clarity. The phonon spectrum in time window $T_2$ shows several interesting features. The satellite peaks BS2 - BS5 have "escaped" the SL due to their finite group velocity and can no longer be detected. This "escape" can be clearly seen in Fig. 1 (c) where the oscillations from BS2 and BS3 are only visible before 680 ps. Based on the acoustic velocities and the SL thicknesses the expected time for BS2 – BS5 modes to leave the SL is 670 ps. The ZC1 mode persists in the spectrum at $T_2$ (Fig. 2 (a)), whilst the ZC2 mode can no longer be detected, hinting at the longer lifetime of ZC1 than ZC2. ZC1 is still present in time window $T_3$.

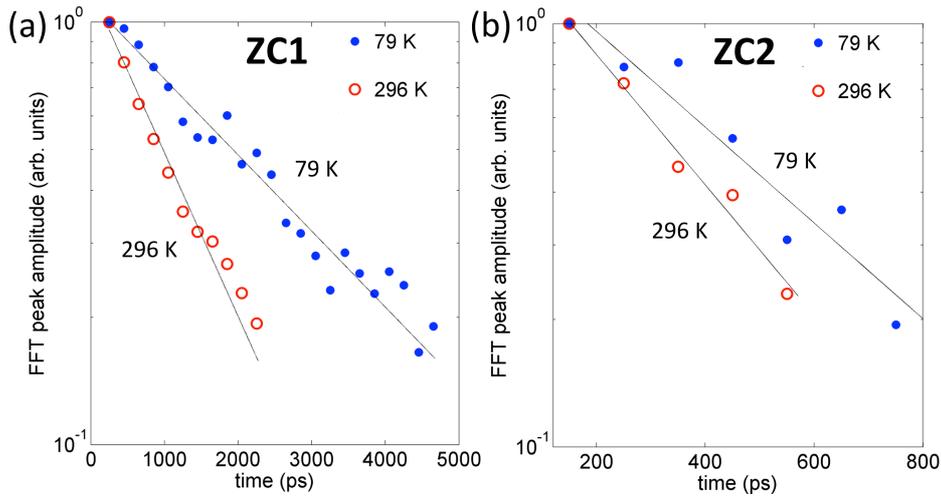

*Figure 3: Time evolution of the sliding Fourier spectrum amplitude for the first zone-center mode ZC1 (0.32 THz) at 296 K and at 79 K (a), and for the second zone-center mode ZC2 (0.64 THz) at 296 K and 79 K (b). Solid lines show exponential fits to the experimental data.*

To assess the lifetime of the ZC1 mode, we monitored the Fourier amplitude within a sliding Hann window [29] with 200 ps FWHM placed at regular 200 ps time intervals. The Fourier amplitude as a function of time for the ZC1 mode at 296 K and 79 K is shown in Fig. 3 (a), normalized to the first measurement point at 260 ps. The data were fitted with a decaying exponential to extract phonon amplitude lifetimes. The same procedure was applied to the ZC2 mode using a Hann window with 100 ps FWHM placed at 100 ps time intervals (Fig. 3 (b)). The repeatability of the decay time measurements was estimated by dividing each measurement set consisting of several hundred scans of the optical delay line into smaller sets of 20 scans. Processing these subsets yielded a spread of lifetimes, the standard deviation of which indicates the repeatability of the measurement. Fig. 3 shows a clear increase in lifetime of the ZC1 mode as temperature is reduced from 296 K to 79 K. We determined amplitude decay times of 1020 ± 40 ps and 2410 ± 120 ps for ZC1 at 296 K and 79 K respectively, corresponding to phonon lifetimes $\tau_{ZC1}(296\ K) = 510 \pm 20$ ps and $\tau_{ZC1}(79\ K) = 1200 \pm 60$ ps. For ZC2 the lifetime increases only a small amount from $\tau_{ZC2}(296\ K) = 150 \pm 20$ ps at *296 K* to $\tau_{ZC2}(79\ K) = 190 \pm 20$ ps at *79 K*. These phonon lifetimes are listed in Table 1.

|  | $\tau\ (296\ K)$ (ps) | $\tau\ (79\ K)$ (ps) | $\tau_{int}(296\ K)$ (ps) |
|---|---|---|---|
| 319.2 GHz (ZC1) | $510 \pm 20$ | $1200 \pm 60$ | $890 \pm 60$ |
| 642.8 GHz (ZC2) | $150 \pm 20$ | $190 \pm 20$ |  |

*Table 1: Measured phonon lifetimes at 296 K and 79 K for phonon modes ZC1 and ZC2 and the computed intrinsic phonon lifetime of ZC1 at 296 K.*

We tried different pump pulse energies from 35 nJ to 360 nJ at both 79 K and 296 K to assess the dependence of phonon lifetime on the number density of excited carriers as mentioned by other authors.[30] In agreement with our previous observations,[24] we could not detect any change as a function of excitation energy. Hence phonon relaxation due to interaction with excited carriers does not appear to play a significant role in our measurements.

Table 2 provides a list of factors contributing to the experimentally measured phonon lifetimes, along with their dependence on frequency and wavelength. Intrinsic (anharmonic) phonon lifetime is due to phonon-phonon scattering caused by anharmonicity of the lattice.[31] At THz frequencies it is described by the Landau-Rumer relaxation model [31 - 33] that applies when $\omega\tau_{th} \gg 1$, where $\tau_{th}$ is the average lifetime of thermal phonons. In the other limiting case, when $\omega\tau_{th} \ll 1$, it follows the Akhiezer model.[31, 34] In both cases the phonon decay rate is thought to scale as $\omega^2$.[31, 35, 36] However at sub-THz frequencies, in the transition region between the two regimes, the exponent is reduced.[19, 24]

Extrinsic phonon scattering is due to scattering from rough interfaces and defects. It is not expected to depend on temperature. In the limit of $L \ll \lambda$, where $L$ is the characteristic size of the defects or roughness, it corresponds to Rayleigh scattering with an $\omega^4$ dependence. On the other hand scattering from surface/interface roughness with $L > \lambda$ follows an $\omega^2$ dependence.[37]

| Process | Frequency dependence | Temperature dependence |
|---|---|---|
| Phonon-phonon scattering (intrinsic) | ≤$\omega^2$ | Yes |
| Interface roughness / defects | $\omega^2$-$\omega^4$ | No |
| Inhomogeneous broadening | $\omega$ | No |
| Finite excitation depth ("run-away" effect) | No | Yes |

*Table 2: Contributors to measured phonon lifetime. (Inhomogeneous broadening and finite excitation depth contribute to the apparent lifetime only.)*

Inhomogeneous broadening refers to the generation of phonons with different frequencies at different locations within the measurement spot due to variations in SL layer thickness. As time progresses phonon oscillations at different locations go out of phase with each other, leading to a decay in coherent signal oscillation amplitude (inhomogeneous dephasing) that masks the true lifetime. Inhomogeneous dephasing is expected to scale linearly with the mode frequency. Indeed, the zone-center mode frequency is given approximately [12] by:

$$f_n = n\left(\frac{d_1}{v_1} + \frac{d_2}{v_2}\right)^{-1}, \qquad (1)$$

where $d_i$ and $v_i$ are layer thicknesses and corresponding acoustic velocities and $n$ is the order number of the zone-center bandgap. A variation $\delta d_1$ in a layer thickness will yield a frequency variation equal to

$$\delta f_n = f_n\left(1 + \frac{v_1 d_2}{v_2 d_1}\right)^{-1} \frac{\delta d_1}{d_1}. \qquad (2)$$

Thus for a given thickness variation in a sample the frequency broadening will be proportional to the frequency of the mode.

For generation of the zone center mode with exactly zero wavevector a spatially uniform excitation is required. Finite penetration depth of the pump beam means that the actual spatial excitation is non-uniform. Thus, rather than generating only at *k=0*, a spread of wave vectors, Δ*k*, is excited in the vicinity of *k=0* for both the ZC1 and the ZC2 modes. This finite width of the excitation in the frequency domain results in an apparent finite lifetime of the excited LA phonons. Another way of looking at this is by recognizing that the excited LA phonons with non-zero *k* have non-zero group velocity and thus "run away" from the excitation/probing region.[38]

In our measurements the attenuation length, *L*, of the pump beam at 296 K was 0.82 μm, estimated using the effective medium approximation. Hence the excited spread of wave vectors is $\Delta k = 1/L = 1.2\ \mu m^{-1} = 5.8 \times 10^{-3}\ (\pi/d)$, where *d* is the SL period. The associated frequency spread of ZC1 and ZC2 modes can be estimated directly from the dispersion curve [24] as $7.7 \times 10^{-2}$ GHz and $5.1 \times 10^{-2}$ GHz respectively, corresponding to decay times greater than 10 ns. This is significantly longer than the measured experimental phonon lifetimes. Hence the contribution of finite penetration depth to phonon lifetime at 296 K is likely to be small. At 79 K the penetration of the pump light is greater, making the effect of finite penetration depth even smaller.

Of the remaining contributors to real and apparent phonon lifetimes, extrinsic scattering and inhomogeneous broadening are independent of temperature. Their contributions to lifetime at 79 K and 296 K are expected to be the same. Only intrinsic lifetime by phonon-phonon scattering is temperature dependent. Recent simulations suggest that below 1 THz, the intrinsic lifetime in GaAs is approximately one order of magnitude larger at 79 K than at 296 K.[40] This is confirmed by our calculations of intrinsic phonon lifetimes in the GaAs/AlAs SL discussed below. The measured lifetime of the ZC1 mode, however, only increased by a factor of 2.4 from 510 ± 20 ps at 296 K to 1200 ± 60 ps at 79 K. This suggests that whilst lifetime at 296 K is due to a combination of intrinsic and extrinsic effects, extrinsic effects must dominate the lifetime at 79 K. The extrinsic, $\tau_{ex}$, and intrinsic, $\tau_{in}$, contributions to total lifetime at 296 K can be combined using Matthiessen's rule:

$$\frac{1}{\tau\,(296\,K)} = \frac{1}{\tau_{in}} + \frac{1}{\tau_{ex}}. \qquad (3)$$

Hence, taking $\tau_{ZC1\,in}^{-1} = \tau_{ZC1}(296\,K)^{-1} - \tau_{ZC1}(79\,K)^{-1}$, the intrinsic lifetime of the ZC1 mode at 296 K can be estimated to be 890 ± 60 ps (listed in Table 1).

For the ZC2 mode there is only a small change in lifetime from 150 ± 20 ps at 296 K to 190 ± 20 ps at 79 K. This suggests that extrinsic effects dominate the 640-GHz phonon lifetimes at both temperatures.

The ZC1 and ZC2 phonon signal decay times at 79 K give a scaling proportional to $\sim\omega^{2.7}$ of the combined extrinsic scattering and inhomogeneous broadening contributions. However, inhomogeneous broadening is expected to scale linearly with frequency. Furthermore, if the decay were dominated by inhomogeneous broadening, it would be expected to be non-exponential. Indeed an inhomogeneous-broadening-dominated decay occurring as oscillators having different frequencies go out of phase typically yields a "semi-bell-shaped" curve with a zero slope at *t=0* such as a Gaussian.[41] Limitations in our experimental signal/noise and dynamic range prevent us from completely ruling out inhomogeneous dephasing, but there are no strong deviations from exponential fits. Thus both the form and the frequency dependence of the observed decay suggest that inhomogeneous dephasing is not a major contributor to our measurements. Hence we conclude that the extrinsic lifetime (measured at 79 K) is dominated by scattering from interface roughness and defects.

**IV. Calculations**

We use a first principles approach to compute intrinsic and extrinsic scattering rates in the SL.[42] The key ingredients necessary for these calculations are the second and third order interatomic force constants. In earlier works [16] these were typically computed using empirical potentials, whilst here DFPT is used. We explicitly take account of interface roughness by modeling it as scattering due to random mass mixing at the interface. Extrinsic scattering rates are then computed using perturbation theory.[43]

The second and third-order interatomic force constants were taken to be the average of the interatomic force constants of pure GaAs and AlAs [44, 45] that were obtained from DFPT.[46] The harmonic force constants were obtained on a 10x10x10 supercell and the anharmonic force constants were found for the nearest neighbors using a supercell approach.[47] For all DFPT calculations, an 8x8x8 Monkhorst-Pack [48] mesh was used to sample electronic states in the Brillouin zone and an energy cutoff of 72 Ry was used for the plane-wave expansion. We carefully tested the convergence of all measured quantities with respect to these parameters. First-principles calculations within density-functional theory were carried out using the PWscf and PHonon codes of the Quantum-ESPRESSO distribution [46] with norm-conserving pseudopotentials based on the approach of von Barth and Car.[49]

The scattering rate, $1/\tau_\lambda$, of a specific phonon mode $\lambda$ is taken to be the sum of a term describing scattering due to interfacial disorder, i.e. extrinsic scattering ($1/\tau_{\lambda\,ex}$), and a term describing anharmonic, i.e. intrinsic, scattering ($1/\tau_{\lambda\,in}$) as in Matthiessen's rule. The anharmonic scattering rates ($1/\tau_{\lambda\,in}$) were computed using the lowest-order three-phonon scattering processes in the single mode relaxation time (SMRT) approximation [50] via:

$$\frac{1}{\tau_{\lambda\,in}} = \pi \sum_{\lambda'\lambda''} |V_3(-\lambda, \lambda', \lambda'')|^2 \times [2(\bar{n}_{\lambda'} - \bar{n}_{\lambda''})\delta(\omega(\lambda) + \omega(\lambda') - \omega(\lambda'')) \\ + (1 + \bar{n}_{\lambda'} + \bar{n}_{\lambda''})\delta(\omega(\lambda) - \omega(\lambda') - \omega(\lambda''))]. \quad (4)$$

Here $V_3(-\lambda, \lambda', \lambda'')$ is the three-phonon coupling matrix or the weighted Fourier transform of the cubic force constants. $\omega$ and $\bar{n}$ are the phonon frequencies and equilibrium phonon populations respectively of the specific phonon mode $\lambda$.

Interface roughness was simulated as a random mixing of Ga and Al atoms in a narrow region on either side of the interface. For short-period SLs this mass-mixing has been previously suggested as the dominant interfacial phonon scattering mechanism.[51] The interfacial scattering rates were computed by replacing the disordered crystal with an ordered crystal and treating the disorder as a perturbation.[43, 52] This approach to computing scattering rates due to mass disorder has been found to yield excellent agreement with experiments.[53] In the SL any atoms at sites affected by the disorder were assigned the average of the Ga and Al masses. All other atoms on either side of the interface were assigned their true masses. We have taken the thickness of the disordered region to be two atomic layers total (i.e. 0.5 nm), consisting of one atomic layer of AlAs and one atomic layer of GaAs with random mixing of Ga and Al atoms. This interface model is informed by HRTEM microscopy studies of interfaces in GaAs/AlAs SLs.[42, 54] The phonon modes of the SL unit cell, with periodic boundary conditions, were used to compute phonon frequencies, group velocities, populations and lifetimes. The scattering rates due to interfacial disorder were calculated using:

$$\frac{1}{\tau_{\lambda\,ex}} = \frac{\pi}{2N} \omega_\lambda^2 \sum_{\lambda'} \delta(\omega_\lambda - \omega_{\lambda'}) \sum_\sigma g(\sigma) |e(\sigma|\lambda')e(\sigma|\lambda)|^2. \qquad (5)$$

where $\sigma$ denotes the atomic sites in the SL unit cell. $N$ is the size used for the discretization of the Brillouin zone. $g(\sigma)$ takes into account the magnitude of the mass disorder and is defined as $g(\sigma) = \sum_i f_i(\sigma)[1 - m_i(\sigma)/\bar{m}_i(\sigma)]^2$, where $i$, $f$ and $m$ denote the atomic species, concentration and mass, respectively. $\bar{m}(\sigma)$ is the average mass at site $\sigma$ and $e(\sigma|\lambda)$ represents the vibration eigenvector for site $\sigma$ and mode $\lambda$. $g(\sigma)$ is non-zero only for atomic sites in the region of disorder.

Anharmonic and interface scattering rates were computed for a number of different SL configurations with different GaAs and AlAs layer thicknesses. The anharmonic (intrinsic) scattering rates for 0.75 / 0.75 nm and 1 / 1 nm SLs, and also for pure GaAs are shown in Fig. 4 (a). Calculations for larger period SLs were not made due to constraints on computational time. The intrinsic scattering rates are largely independent of the SL period. Furthermore they are similar to the scattering rates computed for bulk GaAs, in agreement with previous reports.[16] For comparison with the experimentally studied 14 / 2 nm GaAs/AlAs SL, intrinsic phonon scattering rates were thus taken to be the same as those calculated for the 1 / 1 nm SL. In Fig. 5 fits to the calculated points for the 1 / 1 nm SL intrinsic scattering rates are shown for lattice temperatures of 300 K and 100 K.

The calculated extrinsic scattering rates for 0.75 / 0.75 nm, 1 / 1 nm, 2 / 2 nm and 3 / 1 nm SLs are plotted in Fig. 4 (b). They show approximately linear scaling with the number of interfaces per unit length in the cross plane direction. Extrinsic scattering rates for the 2 / 2 nm and 3 / 1 nm SLs are similar, suggesting that the effect of asymmetrical layer thicknesses is small. Extrinsic scattering rates for comparison with the experimentally measured 14 / 2 nm SL were computed by scaling down the rates found for the 1 / 1 nm SL by a factor of 8, reflecting the difference in the density of interfaces. They are shown in Fig. 5 (dashed line). Also shown on this figure are the experimentally determined intrinsic and extrinsic scattering rates of the ZC1 mode and the extrinsic scattering rate of the ZC2 mode.

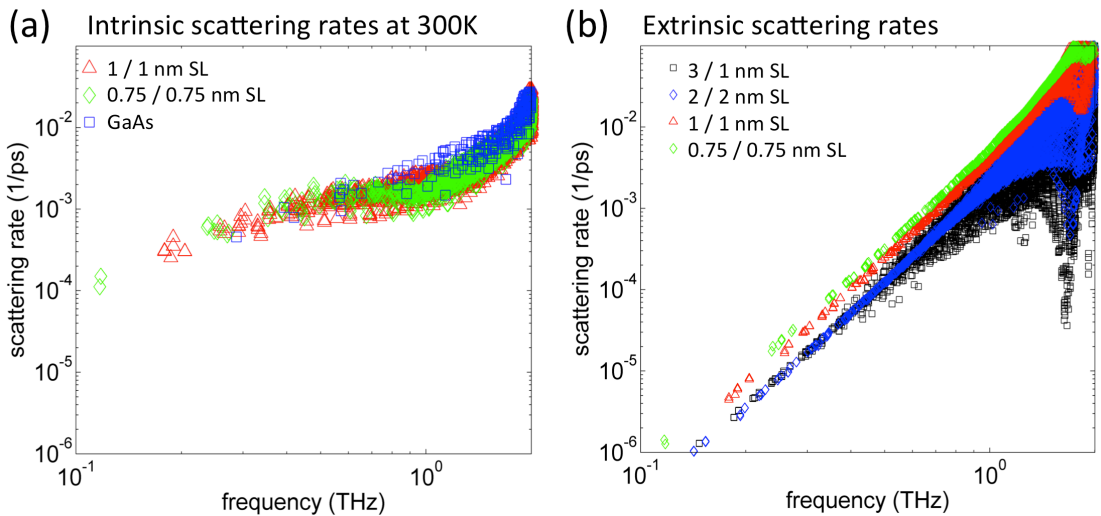

*Figure 4: Raw data from intrinsic and extrinsic scattering rate calculations. a) Intrinsic scattering rates for 1 / 1 nm and 0.75 / 0.75 nm SLs, and bulk GaAs. b) Extrinsic scattering rates for 3 / 1 nm, 2 / 2 nm, 1 / 1 nm and 0.75 / 0.75 nm SLs.*

**V. Discussion**

The measured intrinsic ZC1 mode phonon lifetime of 890 ± 60 ps at 320 GHz and 296 K agrees well with our previous intrinsic lifetime measurement of 950 ± 70 ps at 330 GHz in an 8 nm GaAs / 8 nm AlAs SL.[24] On the other hand extrinsic lifetime at 320 GHz in the present 14 / 2 sample (1200 ± 60 ps) is almost a factor of two longer than that measured at 330 GHz in the 8 / 8 sample (690 ± 30 ps). Indeed, extrinsic scattering is expected to depend on the details of the fabrication process and may vary substantially from sample to sample. The fact that the two samples yielded similar intrinsic lifetimes is in line with the predictions of our model, which yielded similar intrinsic lifetimes for different SLs (see Fig. 4 (a)). The value of the intrinsic lifetime is in relatively good agreement with calculations, as can be seen in Fig. 5. This is interesting since the frequency of 0.3 THz lies at the border of the applicability of the three-phonon scattering model, which requires $\omega \tau_{th} \gg 1$.

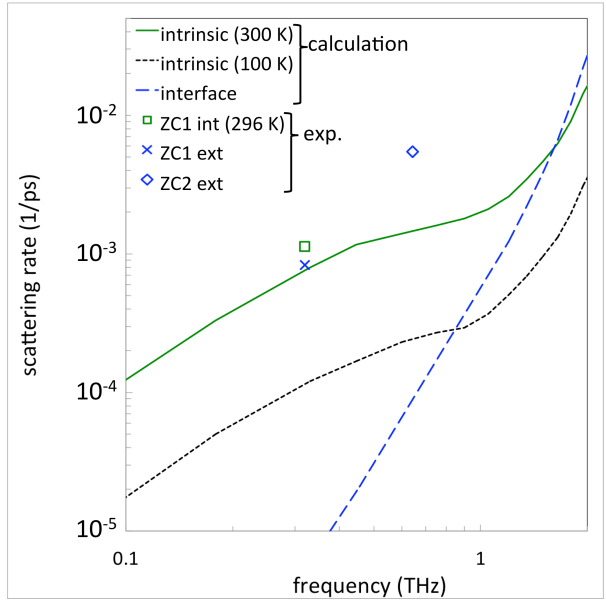

*Figure 5: LA phonon scattering rates: Calculated intrinsic scattering rates for the 14 / 2nm GaAs/AlAs SL at 300 K (solid curve) and 100 K (dotted curve). Calculated interface scattering rate (dashed curve). Symbols show the room-temperature intrinsic (□) and extrinsic scattering rates (✕) of the ZC1 mode, and the extrinsic scattering rate of the ZC2 mode (◇), deduced from the present measurements.*

Whilst the room-temperature phonon lifetime at 320 GHz is determined by both extrinsic and intrinsic contributions, extrinsic effects dominate the lifetime at 640 GHz. This suggests a transition from intrinsically to extrinsically determined phonon lifetime at sub-THz frequencies. The same transition is apparent in the computational results in Fig. 5, although it occurs at a higher frequency. The exact frequency of the crossover of intrinsic to extrinsic phonon scattering rates is highly sensitive to the precise nature of the defects responsible for the extrinsic scattering rate. The computed extrinsic scattering rate in Fig. 5 shows close to $\omega^4$ dependence at sub-THz frequencies. This scaling is expected for Rayleigh scattering [37] and is consistent with the representation of the GaAs/AlAs interfacial roughness in our model as atomic disorder. On the other hand, the measured extrinsic ZC1 and ZC2 scattering rates scale with $\sim \omega^{2.7}$. This reduced exponent suggests that in our measurements scattering from features with larger correlation length plays a role. In fact the scattering from rough interfaces with defect correlation length $L > \lambda$ is expected to scale with $\sim \omega^2$.[37] The observed exponent of *2.7* indicates that we may be in the intermediate regime between Rayleigh and large-scale roughness scattering, closer to the latter. To improve the description of interface scattering a more detailed representation of the interface structure is required in our model, accounting for multiple defect correlation lengths. At present the measurement of two-dimensional interfacial structure is a substantial experimental challenge [54] and it is also likely to vary substantially from sample to sample.

**VI. Summary**

By measuring two coherent phonon modes in a GaAs/AlAs superlattice at 296 K and at 79 K we were able to separate different contributions to phonon lifetimes. At 296 K extrinsic and intrinsic contributions to phonon lifetime at 0.32 THz are about equal, whilst at 0.64 THz extrinsic effects dominate. This indicates a transition from intrinsically to extrinsically dominated phonon lifetime at sub-THz frequencies. The intrinsic lifetime at 0.32 THz is in good agreement with our previous measurements and calculations. On the other hand, the calculations of extrinsic scattering rates, based on the interfacial atomic disorder model, yield values lower than the measured rates and do not match the observed frequency dependence. An accurate model of phonon scattering by real interfaces based on quantitative characterization of the interface roughness is a challenge lying ahead.

**Acknowledgments**

FH would like to acknowledge funding by the U.S. Department of Energy, Office of Basic Energy Sciences under Award Number: DE-FG02-00ER15087. Furthermore this work was supported as part of the "Solid State Solar-Thermal Energy Conversion Center (S3TEC)", an Energy Frontier Research Center funded by the U.S. Department of Energy, Office of Basic Energy Sciences under Award Number: DE-SC0001299/DE-FG02-09ER46577.

[38] It is instructive to consider a situation when the medium is transparent at the probe wavelength and the measurement is done in transmission (forward scattering) geometry. In this case the observed phonons never run out of the probed region. Indeed, forward-scattering geometry is selectively sensitive to k=0 phonons.[39] Thus even if the finite penetration depth of excitation light yields a spread of wavevectors, the detection will be selectively sensitive to the zone-center mode, and neither frequency broadening nor associated signal decay will be observed.